\documentstyle[prc,aps,floats,psfig]{revtex}
\begin{document}
\draft

\title{Comment on ``Determination of the chiral coupling constants $c_3$ and
$c_4$ in new $pp$ and $np$ partial-wave analyses''}

\author{D. R. Entem$^{1,2,}$\footnote{Electronic addresses: 
        dentem@uidaho.edu, entem@usal.es}
        and R. Machleidt$^{1,}$\footnote{Electronic address: machleid@uidaho.edu}}

\address{$^1$Department of Physics, University of Idaho, Moscow, ID 83844, USA\\
  $^2$Nuclear Physics Group, University of Salamanca, E-37008 Salamanca, Spain}

\date{\today}

\maketitle

\begin{abstract}
In a recent study [M.C.M. Rentmeester {\it et al.}, 
Phys.\ Rev.\ C {\bf 67}, 044001 (2003)], the Nijmegen group reports on
the determination of the chiral low-energy constants (LEC), $c_3$ and $c_4$,
involved in the 2$\pi$-exchange part of the $NN$ amplitude at 
next-to-next-to-leading order (NNLO) of chiral perturbation theory.
This analysis does not apply 
the uniquely-determined and model-independent $NN$ amplitudes
at NNLO and uses, instead, amplitudes that are up 90\% smaller.
We point out that this flaw produces a large systematic error,
rendering the
Nijmegen method unsuitable for a reliable determination of the LEC.
\end{abstract}


In a recent paper~\cite{RTS03}, the Nijmegen group reports on a study
of the chiral two-pion exchange interaction 
which contains the low-energy constants (LEC) $c_1$, $c_3$, and $c_4$.
The long-range part of this interaction is applied in an analysis
of the world $pp$ and $np$ data below 500 MeV laboratory energy.
The authors state that, based upon this analysis,
they were able to determine the LEC
$c_3$ and $c_4$ with an accuracy of 2-5\%.
This recent Nijmegen analysis~\cite{RTS03} is an update of
an earlier one~\cite{Ren99}, and the present note applies
to both.

It is the purpose of this Comment to point out that
the Nijmegen determination of the LEC appears to have
basic
flaws which can be estimated to cause a large systematic error.

The traditional approach for describing nuclear forces
has been the meson-exchange model in which mesons of
increasing mass are used to generate contributions of
decreasing range. This expansion is then truncated
at a certain range which is believed to be unimportant
for traditional nuclear physics purposes.

The new approach to nuclear forces, that was initiated by
Weinberg \cite{Wei90} and pioneered
by Ord\'o\~nez~\cite{OK92}, Ray, and van Kolck~\cite{ORK94,Kol99},
has a foundation and philosophy which are quite different
from meson phenomenology. Based upon the chiral symmetry
of QCD, an expansion is made in powers of the nucleon momenta.
This is known as chiral perturbation theory ($\chi$PT).

The crucial point to realize is that---in contrast to 
meson phenomenology---$\chi$PT is a {\it theory} rather than a
model, i.~e., it makes 
{\it exact} predictions at each order (if the constants of the
theory are known).

Physical constants are to be determined in a
model-independent way, if by all means possible.
While model-independent determinations are {\it per se\/} impossible
if one is dealing with a model or phenomenology,
model-independence is inherent to the framework of $\chi$PT.

At next-to-next-to-leading order (NNLO), the theory consists
of one-pion-exchange (OPE), two-pion-exchange (TPE)
at ${\cal O}(Q^3)$~\cite{KBW97}, 
where $Q$ denotes a momentum or pion
mass, and contact terms of order $Q^0$ and $Q^2$. These
contributions define {\it uniquely\/} the $NN$ amplitude
at NNLO. A problem is that the parameters of the contact
terms are not known. However, since these contacts
contribute only to $S$ and $P$ wave amplitudes,
the $NN$ amplitudes for partial waves with orbital
angular momentum $L\geq 2$ do not receive
any contact contributions~\cite{EGM98}.
Consequently, at NNLO, the $NN$ amplitudes for $D$ and
higher partial waves are determined
by OPE and TPE at ${\cal O}(Q^3)$
in an unique and completely model-independent way~\cite{KBW97,EGM98}.
These two contributions depend 
(apart from the nucleon and pion mass)
on five constants:
the axial-vector coupling constant, $g_A$,
the pion decay constant, $f_\pi$,
and three LEC that appear in the dimension-two $\pi N$
Lagrangian and are, conventionally, denoted by
$c_1$, $c_3$, and $c_4$.
The constants $g_A$ and $f_\pi$ are known from other sources
(we use $g_A=1.29$ and $f_\pi=92.4$ MeV and the same is used
in Ref.~\cite{RTS03}) and for $c_1$ the educated estimate
$c_1=-0.76$ GeV$^{-1}$ can be made~\cite{RTS03,Ren99}.
Now only two constants are open, namely, $c_3$ and $c_4$.
Since the LEC are also involved in the $\pi N$ amplitude,
these constants have been determined by calculating
the $\pi N$ amplitude up to a certain order of $\chi$PT
and adjusting the constants such that
the empirical $\pi N$ information is fit~\cite{FMS98,BM00}.
Similarly, the
$NN$ amplitudes of partial waves with $L\geq 2$
can be compared with the corresponding empirical
amplitudes to extract $c_3$ and $c_4$.

Once all five constants are fixed, then there exists
a unique prediction for all $NN$ amplitudes with $L\geq 2$.
Using $c_3=-4.78$ GeV$^{-1}$ and $c_4=3.96$ GeV$^{-1}$,
these uniquely determined $NN$ amplitudes are shown
in Fig.~1 by the solid lines for one $D$ and one
$F$ wave case (see Refs.~\cite{KBW97,EM02} for the
details of how these amplitudes are calculated
perturbatively; conventions as in Ref.~\cite{EM02}).

{\it It must be stressed again that we are dealing
here with a theory and that a theory makes
exact predictions. Thus, the solid line in Fig.~1 is 
the unique result of $\chi$PT at NNLO for the NN system
(using the above constants).
Using the same constants, there is no other result
at NNLO and every theoretical physicist will independently
reproduce this result.}

However, when we compare this result with the one the Nijmegen
group obtains and uses in their analysis (dashed line in Fig.~1), 
we notice a discrepancy of up to 90\%.
When such amplitudes are applied in an analysis,
a large systematic
error is produced and the extracted constants
will carry such large error. 

The Nijmegen amplitudes are obtained as a result
of representing
the chiral OPE and TPE at NNLO
in terms of an $r$-space potential that is cut off
at $r=1.6$ fm. 
The motivation for doing this may be a confusion between
the old range argument (that applies to meson-model
based potentials) and the fundamentals of
$\chi$PT which is an expansion in powers of nucleon
momenta, and not in terms of ranges.
As pointed out before,
$\chi$PT has, by principle, no model-dependence,
whereas model-dependence is clearly introduced
in the analyses of Refs.~\cite{RTS03,Ren99}.

Another point of concern is that the Nijmegen group uses
$\chi$PT at NNLO up to
500 MeV. It is wellknown that $\chi$PT at NNLO
is appropriate only up to 50 MeV in $D$ waves and
about 150 MeV in $F$ waves~\cite{KBW97,EM02}.

In conclusion,
the method of determination of the chiral LEC, $c_3$ and $c_4$,
applied in the recent as well as the earlier Nijmegen 
analyses~\cite{RTS03,Ren99} may be unreliable.

This work was supported in part by the U.S. National Science
Foundation under Grant No.~PHY-0099444,
the Spanish Ministerio de Ciencia y Tecnolog{\'\i}a 
under Contract No. BFM2001-3563, and the Spanish Junta de Castilla y Le\'on under 
Contract No. SA-109/01.

\begin{figure}
\vspace{-5cm}
\hspace{2cm}
\psfig{figure=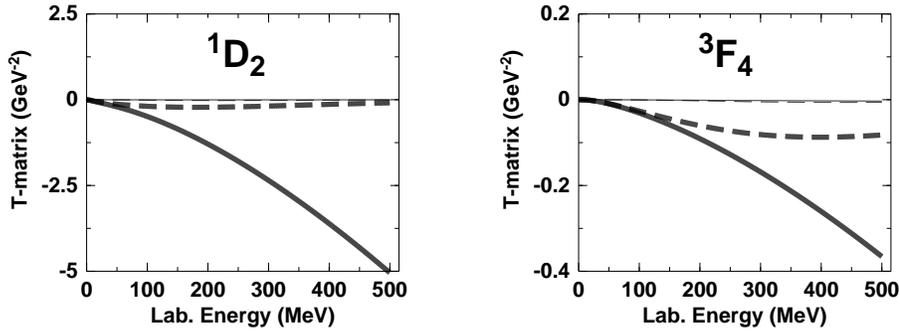,height=16.0cm}
\vspace{-6.5cm}
\caption{On-shell $NN$ amplitudes 
up to 500 MeV laboratory energy
for the $^1D_2$ and $^3F_4$ states.
The solid lines represent the correct amplitudes 
for OPE plus TPE at NNLO (using the constants given in the text).
The dashed lines show the amplitudes used in the Nijmegen
analysis for OPE plus TPE at NNLO (using the same constants).
Thick lines represent the real parts of the amplitudes
and thin lines the imaginary parts. (Note that the thin lines are $\approx 0$
on the scale of the figures
and hardly distinguishable.)}
\end{figure}


\begin{references}
\bibitem{RTS03} M. C. M. Rentmeester, R. G. E. Timmermans,
and J. J. de Swart, 
Phys.~Rev.~C {\bf 67}, 044001 (2003).
\bibitem{Ren99} M. C. M. Rentmeester, R. G. E. Timmermans,
J. L. Friar, and J. J. de Swart, 
Phys. Rev. Lett. {\bf 82}, 4992 (1999).
\bibitem{Wei90} S. Weinberg, Phys.\ Lett.\ B {\bf 251}, 288 (1990);
Nucl.\ Phys.\ {\bf B363}, 3 (1991).
\bibitem{OK92}
C. Ord\'o\~nez and U. van Kolck,
Phys.\ Lett.\ B {\bf 291}, 459 (1992).
\bibitem{ORK94}
C. Ord\'o\~nez, L. Ray, and U. van Kolck,
Phys.\ Rev.\ Lett.\ {\bf 72}, 1982 (1994);
Phys.\ Rev.\ C {\bf 53}, 2086 (1996).
\bibitem{Kol99} U. van Kolck, Prog.\ Part.\ Nucl.\ Phys.\ {\bf 43}, 337 (1999).
\bibitem{KBW97} N. Kaiser, R. Brockmann, and W. Weise,
Nucl.\ Phys.\ {\bf A625}, 758 (1997).
\bibitem{EGM98} E. Epelbaum, W. Gl\"ockle, and U.-G. Mei\ss ner,
Nucl.\ Phys.\ {\bf A671}, 295 (2000).
\bibitem{FMS98} N. Fettes, U.-G. Mei\ss ner, S. Steiniger,
Nucl.\ Phys.\ {\bf A640}, 199 (1998).
\bibitem{BM00} P. B\"{u}ttiker and U.-G. Mei\ss ner,
Nucl.\ Phys.\ {\bf A668}, 97 (2000).
\bibitem{EM02} D. R. Entem and R. Machleidt, Phys. Rev. C {\bf 66}, 014002 (2002).
\end{references}
\end{document}